# Explainable Machine Learning in Healthcare: Methods, Interpretation, and Applications for Clinical Research

*Krishna Padmanabhan, PhD[1]†, Minxin Lu, PhD[2]†*, Dai Feng, PhD[3], Natalia Kan-Dobrosky, PhD[4], Sai Konduri, MD[5], Heather J. Litman, PhD[6], Achilleas Livieratos, PhD[7]*

[1] Biostatistics, R&D, Madrigal Pharmaceuticals Inc, Conshohocken, PA, 19428, USA
[2] Department of Nephrology, Boston University School of Medicine, Boston, MA, 02118, USA
[3] MA&HTA Statistics, Data and Statistical Sciences, AbbVie Inc., North Chicago, IL, 60064, USA
[4] Statistical Science, Data and Statistical Sciences, AbbVie Inc., North Chicago, IL, 60064, USA
[5] Pulmonary-Critical Care, St. Elizabeth Healthcare, Fort Mitchell, KY, 41017, USA
[6] Biostatistics, PPD CorEvitas Clinical Registries, Thermo Fisher Scientific, Waltham, MA, 02451, USA
[7] Independent Researcher, Athens, 152 38, Greece

† These authors contributed equally as co–first authors.

*Corresponding author:

**Minxin Lu, PhD**

650 Albany St, Floor 5, Boston, MA, 02118

Email: minxinlu@bu.edu

# Abstract

**Objectives**: To provide a practical and methodologically grounded overview of explainable machine learning (XML) approaches in healthcare, with emphasis on their interpretation and application in clinical research and decision support. By moving beyond traditional predictive models, this primer aims to foster trust, transparency, and informed clinical decision-making, ultimately bridging the gap between data science and medical practice.

**Materials and Methods**: We present a structured review of commonly used XML methodologies, including global and local interpretability tools such as SHapley Additive exPlanations (SHAP), Local Interpretable Model-Agnostic Explanations (LIME), Partial Dependence Plots (PDP), and Individual Conditional Expectation (ICE) plots. For each method, we explain the underlying mechanism at a high level, visualize representative outputs, and provide structured guidance on interpretation, appropriate use, and limitations, illustrated using the publicly available Heart Disease dataset.

**Results**: XML techniques provided intuitive visual and quantitative insights into how predictors influence model predictions. Global methods characterized population-level feature effects, whereas local methods revealed patient-level contributions useful for individualized interpretation. Our worked examples demonstrate how XML outputs can identify nonlinear relationships, detect interaction effects, and reveal heterogeneity in predicted risk across patients, addressing key challenges in translating ML predictions into interpretable outputs for clinical research.

**Discussion and Conclusion**: XML tools offer valuable interpretability for ML models and support more transparent and accountable ML applications in clinical research. By providing a methodologically grounded overview alongside practical implementation examples and structured guidance on each method's strengths and limitations, this primer helps bridge the gap between advanced ML methodology and clinical applicability. Thoughtful adoption of XML approaches may facilitate better understanding, communication, and critical evaluation of ML predictions in healthcare research, ultimately supporting evidence-based clinical decision-making.

# 1. Introduction

The integration of Machine learning (ML) into healthcare is transforming how biomedical informaticians, clinical investigators, clinicians, and healthcare data scientists process data, diagnose conditions, and personalize treatments [1]. ML offers a powerful tool for synthesizing multidimensional datasets, identifying patterns, and predicting outcomes at a large scale [2]. However, the clinical value of ML tools depends on practitioners' ability to interpret outputs within biological and pathophysiological contexts. For example, sepsis risk forecast must be reconciled with bedside observations and patient history to avoid alarm fatigue [3]. Despite strong interest, 84% of physicians find it important to receive "proper training/education on AI tools being used" for the adoption of AI into their practice, only 12% of clinicians report using them for assistive diagnostics [4]. Proficiency in explainability techniques is therefore critical for biomedical scientists to translate predictions into interpretable outputs and communicable model behaviors that support clinical judgements.

## 1.1 Explainable Machine Learning (XML)

XML focuses on making the ML decision-making more understandable [5–12], analogous to how domain experts translate complex medical knowledge into comprehensible explanations for patients [13]. XML covers three related concepts: transparency, interpretability, and explainability [10]. Transparency focuses on the model's internal processes [10], including model structure, training data, population coverage, and performance characteristics. Interpretability addresses how a model works [5] and refers to the overall model's behavior and logic [13,14]. Explainability provides the rationale behind an individual prediction and answers "what is the model saying?" [5,13].

## 1.2 Importance of XML

XML was introduced to address "black-box" models, such as deep learning systems with millions of parameters, that are difficult to interpret in clinical contexts [6,15–18]. The risk of deploying non-explainable ML systems has been demonstrated by several real-world examples: Zech et al. (2018) [19] proved that a deep learning model for pneumonia detection relied on hospital-specific artifacts on the chest X-rays rather than actual pathology; Caruana et al. (2015) [20] demonstrated that a black-box risk model incorrectly learned that asthma patients had lower complication risk, because it failed to account for the aggressive care these patients typically receive; and Adamson & Smith. (2018) [21] showed that an algorithm trained predominantly on Western populations can perform poorly in underrepresented groups and exacerbate health disparities. These examples underscore the importance of interpretability tools to make ML systems more understandable [22].

### 1.3 Regulatory Perspectives

The integration of AI/ML into healthcare has promoted global regulatory agencies to establish frameworks to ensure the use of ML is safe, transparent, and accountable. In 2021, the guiding principles for Good ML Practice was jointly released by the U.S. Food and Drug Administration (FDA), Health Canada, and the United Kingdom's Medicines and Healthcare products Regulatory Agency (MHRA) [23]. In the same year, FDA also released an Action Plan for AI/ML-Based Software as a Medical Device (SaMD) [24]. The EU Artificial Intelligence Act [25], introduced requirement for AI transparency, mandating AI providers to disclose their use, purpose, and limitations to support informed adoption and accountability. In January 2025, the FDA published a draft guidance outlining a seven-step framework for evaluating the credibility of AI models [26].

## 1.4 Conclusion

Biomedical scientists must be able to evaluate models confidently, explain the model's output [22], identify bias in training data, ensure fairness of model process, and confirm that a model behaves as expected [8], all of which are essential to developing safer and more accountable applications [27]. While prior reviews have addressed XML methods broadly (e.g., Angelov et al., 2021[6], Dwivedi et al., 2023 [28]; Molnar, 2022 [11], Carvalho, 2019 [12]) or within specific clinical domains (e.g., Petch et al., 2022 [13]), our work offers a method-oriented primer aimed at a broader informatics audience with structured guidance on interpretation and limitations of each method. We use the term XML to denote explainable machine learning techniques to reflect common usage rather than technical accuracy.

## 2. Overview

### 2.1 Overview of XML Methods

**Table 1**

Classification of XML methods.

| Type | Method Type | Examples | Notes |
|---|---|---|---|
| Post-hoc | Model-specific | - Feature Importance (for Tree-Based Models)<br>- Attention maps (for Neural Networks)<br>-Attention weights (for Transformers)<br>- Layer-wise relevance propagation (LRP) | Tailored to a specific model architecture |
| Post-hoc | Model-agnostic | - SHapley Additive exPlanations (SHAP) [31]<br>- Local Interpretable Model-Agnostic Explanations (LIME) [22]<br>-Partial dependence plots (PDPs) [32]<br>- Individual Conditional Expectation (ICE) plots [33] | Work with any model after training. They probe the black-box to understand its feature effects and decision behaviors. |
| Intrinsic | Interpretable Models | - Linear/logistic regression<br>- Decision trees<br>- Rule-based models such as RuleFit [34], Bayesian Rule Lists [35] | Provide transparency by design, similar to clinical scoring systems. |

The XML methods detailed in this primer were chosen to maximize clinical research utility based on three criteria. First, we aimed for broad applicability by prioritizing model-agnostic approaches (SHAP, LIME, PDP, ICE) that provide consistent explanations across algorithms, while also including established model-specific measures (Gini importance, regression coefficients) that remain foundational in clinical research. Second, we focused on tabular clinical data, the most common format for risk prediction in clinical informatics; accordingly, modality-specific techniques such as class activation maps (CAM) and layer-wise relevance propagation (LRP), essential for deep learning in medical imaging [29,30], were excluded to maintain a focused scope. Finally, we selected methods that balance global explanations of population-level predictors with local explanations of individual patient risk. These XML methods can be broadly categorized based on when and how they provide explanations into three categories (Table 1).

### 2.2 Dataset for Demonstrating XML Methods

The utility of XML is best demonstrated by looking at a dataset that reflects clinical complexity. We select the Heart Disease Dataset [36] for its diversity and real-world relevance. It contains 1,189 samples drawn from five hospital systems with 11 clinically interpretable features, including demographic, physiological, and diagnostic variables, that resemble structured clinical data commonly used in predictive modeling. This structure allows explanation methods to be illustrated in a transparent and easily interpretable manner for biomedical scientists in cardiology risk assessment models. While this example focuses on tabular clinical variables, the XML techniques are model-agnostic and broadly applicable to other data modalities used in healthcare machine learning, including imaging, genomic data, and high-dimensional electronic health record (EHR) features. Because the models used here are purely predictive, the explanations describe how the model uses features to generate predictions. These explanations should be interpreted as describing associations learned by the model rather than causal relationships between variables and outcomes. Table 2 presents descriptions and brief explanations for all variables in the dataset. The analysis is performed on R version 4.5.1.

**Table 2**

Heart Disease Data Description of Attributes (Based on Siddhartha, 2024).

| Attribute | Description | Brief Explanation |
|---|---|---|
| Age | Measured in years | Risk of coronary artery disease increases with age due to development of atherosclerosis, one of the primary pathophysiological processes of coronary artery disease [37] |
| Sex | Value 1: Male<br>Value 0: Female | Risk of coronary artery disease is higher in men compared to women [37]. |
| Chest pain type | Value 1: Typical angina<br>Value 2: Atypical angina<br>Value 3: Non-anginal pain<br>Value 4: Asymptomatic | Classically, “typical” angina is a symptom associated with coronary artery disease. However, atypical or absence of angina may also be a feature of the disease. |
| Serum Cholesterol | In mg/dl | Higher cholesterol generally increases plaque formation in coronary arteries, increasing the risk of atherosclerosis. Although further subsets of cholesterol have more specific correlation with coronary artery disease, the author of the dataset has not included the information [37]. |
| Fast Blood Sugar | Value 1: Fasting blood sugar > 120 mg/dl<br>Value 0: Fasting blood sugar <= 120 mg/dl | Elevated fasting blood sugar (glucose) may suggest the risk of diabetes mellitus, another major risk factor for coronary artery disease [37]. |
| Resting electrocardiogram Results | Value 0: Normal<br>Value 1: Having ST-T wave abnormality (T wave inversions and/or ST elevation or depression of > 0.05 mV)<br>Value 2: Showing probable or definite left ventricular hypertrophy by Estes' criteria | ST-T wave abnormalities may indicate myocardial ischemia or prior myocardial injury. Left ventricular hypertrophy suggests chronic hypertension, another risk factor for coronary artery disease [37]. |

| Maximum heart rate achieved | 71–202 | Lower exercise heart-rate capacity may suggest impaired cardiovascular fitness. |
|---|---|---|
| Exercise induced angina | Value 1:Yes<br>Value 0: No | Angina during exercise is a strong predictor for coronary artery disease. |
| Oldpeak | Depression | A higher oldpeak value (greater ST depression) reflects exercise-induced myocardial ischemia which correlates with coronary artery disease [38]. |
| ST slope | Value 1: Upsloping<br>Value 2: Flat<br>Value 3: Downsloping | ST segments changes show strong association with myocardial ischemia [38]. |
| Coronary artery disease | 1 = Heart disease, 0 = Normal | Indicates the confirmed presence or absence of coronary artery disease. This is our outcome variable. |

## 3. Application of XML Methods

The following sections outline representative XML methods demonstrated using a Random Forest prediction model on the Heart Disease dataset. Since the focus is on explainability, we have not performed test-train split or optimization of the ML algorithm itself. For each method, we explain the mechanism at a high level, visualize outputs, and provide interpretation, limitations, and guidance on appropriate use.

Before reviewing individual methods, it is useful to consider what clinical questions each method answers. Table 3 maps common questions a biomedical scientist or clinical researcher may have about an ML model to the XML methods that address them. The sections that follow are organized by method type, but readers may also use this table to navigate directly to the method most relevant to their question.

**Table 3**

Overview of XML Methods by Clinical Questions.

| Clinical Question | XML Method(s) | Section |
|---|---|---|
| Which variables are most important to the model's predictions? | Feature importance (Gini, permutation), SHAP Beeswarm plot | 3.1, 3.2.1 |
| Why did this specific patient receive this prediction? | SHAP waterfall plot, LIME | 3.2.1, 3.2.2 |
| How does changing one variable affect predicted risk across patients? | PDP, ICE plots | 3.2.3, 3.2.4 |
| Can I get human-readable decision rules from the model? | RuleFit, Bayesian Rule Lists | 3.3.2 |

### 3.1. Model-Specific Methods

Model-specific explanation methods take advantage of the structure and parameters of a particular model type [28]. They "open up the black box" and use model internals, such as coefficients, tree splits, or neural network weights, to provide an explanation.

#### Feature Importance in Tree-Based Models

A common first question about any predictive model results is which variables matter most. In linear models, standardized coefficients offer a straightforward measure of feature importance. Decision trees [39] and tree-based ensembles, such as random forest [40], gradient boosting [32] and XGBoost [41], capture complex non-linear relationships. Feature importance in tree models is generally defined by the reduction in prediction error or impurity when a variable is used for splitting. In random forest, the impurity-based feature importance is calculated by summing Gini impurity, or other split criteria, reduction across all trees, whenever that feature is used to split a node. Features that yield larger reductions in impurity (increase in model accuracy) receive higher importance scores. An alternative model-agnostic measure is permutation importance, which measures how randomizing a feature's values drops the model's accuracy. The impurity-based importance is more commonly used for tree models since it comes directly from the model construction.

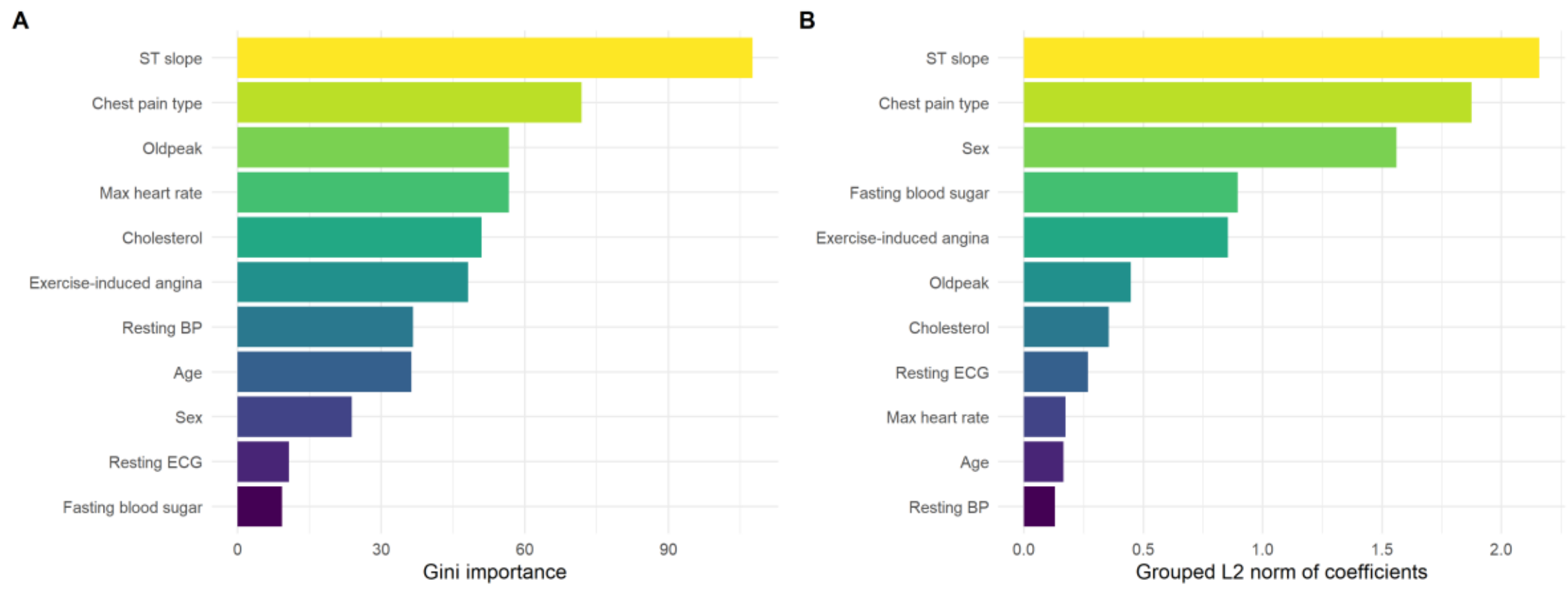

**Fig. 1.** (A) Random Forest Gini Importance (B) Logistic Regression Variable importance (grouped L2 norm of coefficients)

**Interpretation**: In this coronary artery disease prediction (Fig. 1), Random Forest (left) and Logistic Regression (right) rank features by predictive importance. Bar length indicates each feature's contribution. The x-axis shows an importance score, while the y-axis lists the variables ranked from most to least important (from top to bottom). For logistic regression, standardized coefficients serve as covariate importance measures and can be interpreted as odds ratios to assess feature effects. Both models assign high importance to features fundamental to cardiac evaluation, such as ST slope and chest pain characteristics. This convergence across the two methodological distinct methods suggests internal consistency of highlighted important variables within the dataset. The variation in rankings reflects the differences in ways these models encode feature relationships. The Gini importance of Random Forest captures non-linear relationships such as threshold effects and interaction effects. In contrast, the coefficients of Logistic Regression describe linear, monotonic, independent contributions to the log-odds. It should be noted that convergence in feature importance does not indicate predictive performance, generalizability to new patients requires evaluation with performance metrics on a held-out test set.

**When not to use:** Model-specific methods should not be used when explanations need to be compared across different models, as they only apply to a single model type. They can also be misleading if the model is biased or misspecified, since the explanations will reflect those issues.

**Limitations**: Feature importance indicates which variables contribute most to the model's predictions, but it does not indicate the direction of association (whether higher values increase or decrease predicted risk), threshold effects (at what values features become influential), or interactions between variables. These insights require complementary tools such as partial dependence plots or SHAP values.

### 3.2. Model-Agnostic Methods

Model-agnostic methods work for any ML model and do not require knowing the internal structure or weights. They usually work by perturbing inputs, observing outputs, and computing relationships. Important model-agnostic techniques include SHAP and LIME for local (patient level) explanation and Partial Dependence and ICE plots for global (overall) explanation for feature effect pattern visualization.

#### 3.2.1. SHAP (SHapley Additive exPlanations)

**What it is:** Clinical investigators often need to know which factors drove a specific patient's risk estimate and by how much. SHAP [31] is a widely-used method and often considered the "gold standard" for explaining individual predictions by quantifying each feature's contribution to the total deviation from the population average. SHAP feature contributions are additive and sum precisely to the prediction difference. For example, if the average risk is 10% and a patient's risk is 15%, SHAP might show: Age (+2%), Blood pressure (+4%), Current medications (-1%) = +5%.

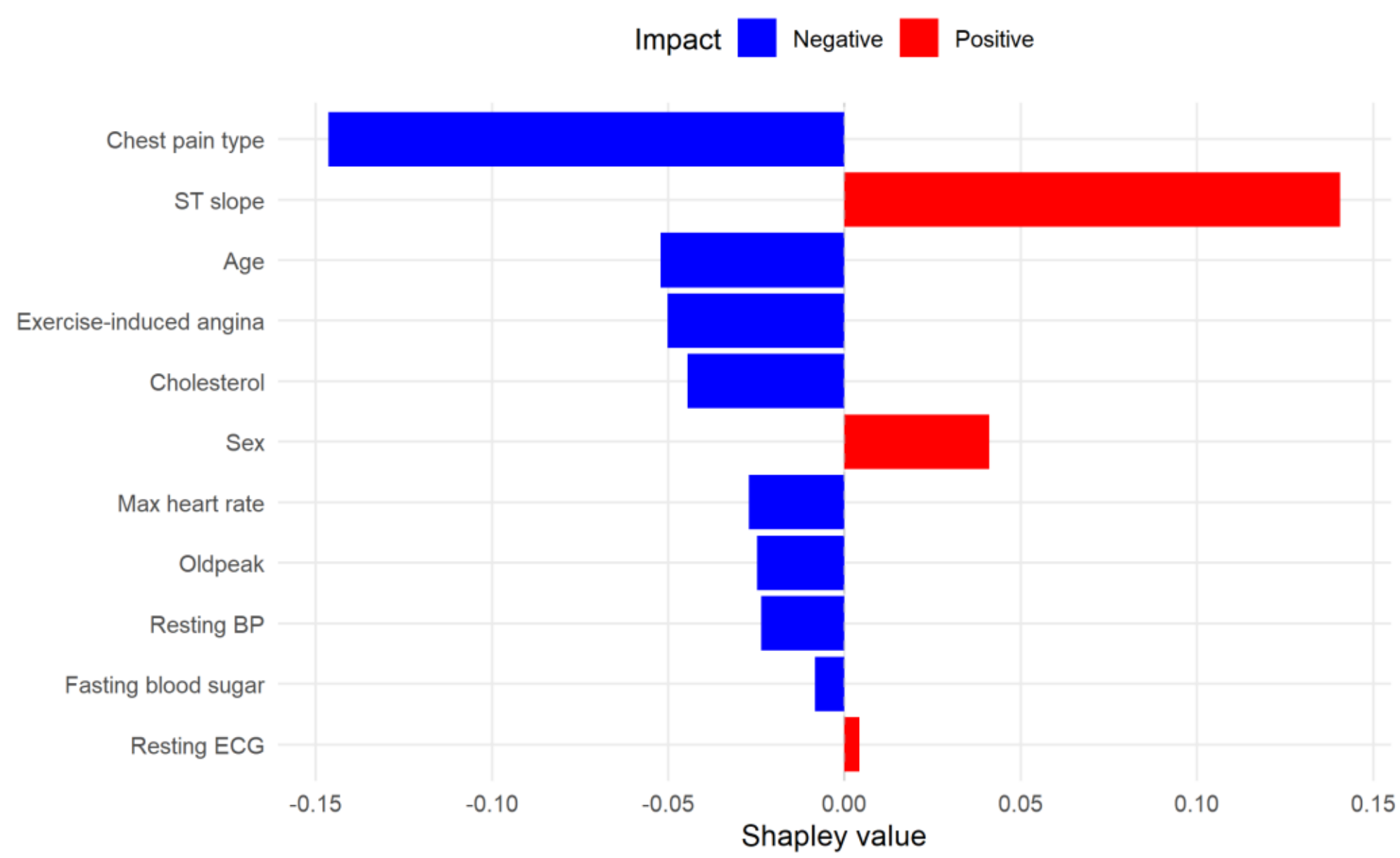

**Fig. 2.** SHAP (Shapley) waterfall plot – patient 22.

**Interpretation**: The SHAP waterfall plot (Fig. 2) for a randomly selected Patient 22 shows how the Random Forest model's prediction was constituted by features that decreased (blue) and increased (red) the predicted risk score. The largest negative SHAP contribution was chest pain type (-0.15), partially offset by ST slope (+0.14) and other factors. The sum of the SHAP values is -0.1915, which indicates a ~36.37% lower predicted risk compared to the "average" patient. The SHAP values represent the contribution of each feature to the model's prediction on the probability scale. This plot may also be displayed on the log-odds scale.

**When to use:** SHAP provides theoretically grounded patient-level insights by explicitly quantifying each feature's contribution to the model prediction. This is particularly useful for cases with mixed risk profiles, where SHAP clarifies which variables most influenced the model's decision.

**When not to use:** SHAP should not be used to compare individual patients across subgroups, as attributions are not calibrated across different patient populations. SHAP should not be used to show the general relationship between a variable and predicted risk. This is better accomplished by Partial Dependence Plots (PDP). Moreover, the exponential computational complexity of standard SHAP makes it impractical for large high-dimensional datasets without optimization. LIME is often a more efficient choice for these scenarios.

**Limitations:** SHAP assumes that features contribute independently to predictions, which may not reflect true biological interactions. SHAP values capture the model associations, not necessarily causation; a high SHAP value could signal an unmeasured proxy, not a direct cause. Furthermore, the "average" patient in the data may not represent typical clinical cases. Additionally, SHAP explanations can be unstable for patients near decision boundaries.

**Global Feature Contribution Patterns via SHAP**

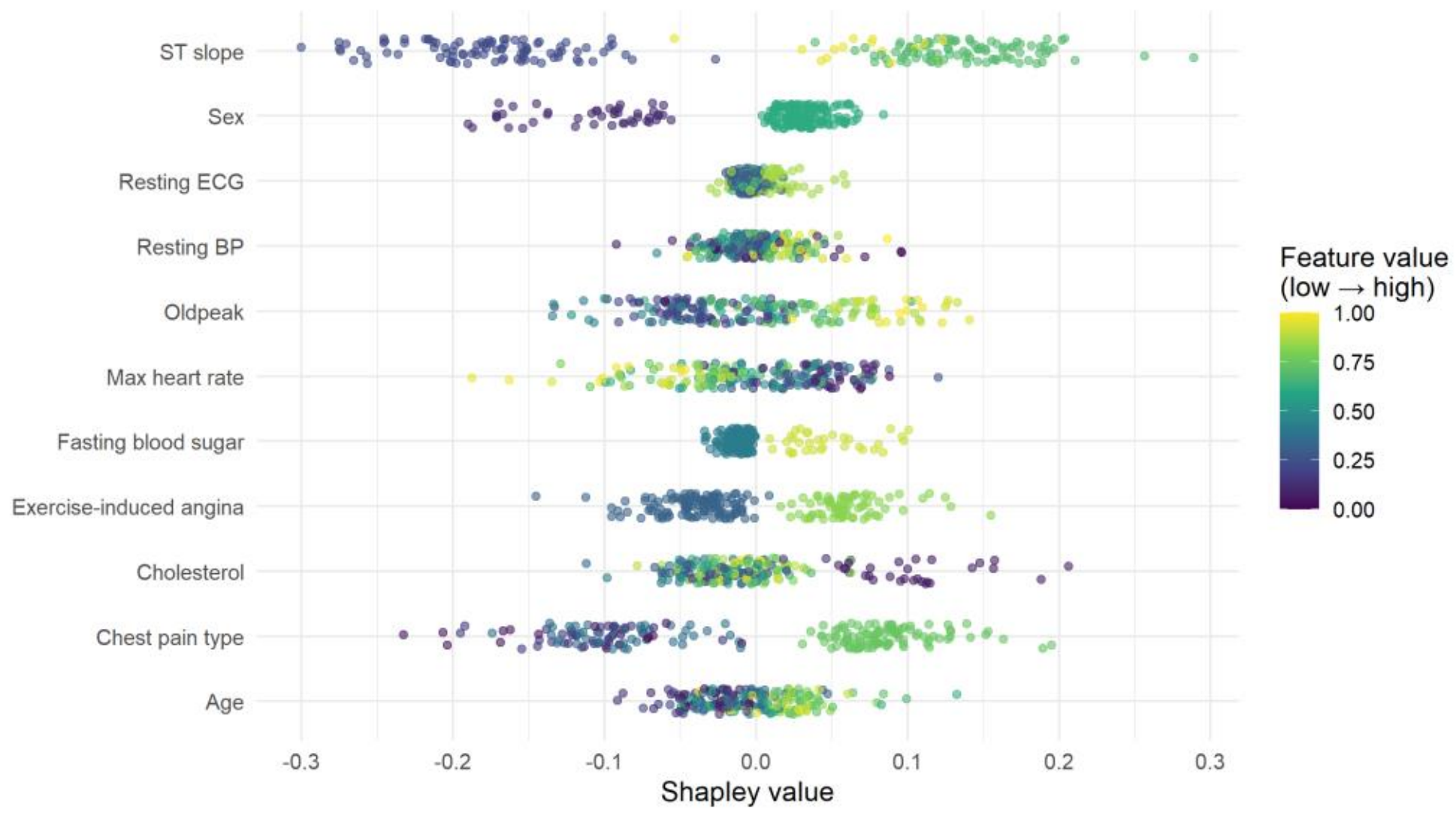

**Fig. 3.** SHAP Beeswarm plot, colored by feature value. Shapley values show the impact on probability.

**Interpretation**: SHAP Beeswarm plot (Fig. 3) aggregates SHAP values to reveal population-level patterns of feature impact distributions, directions, and interactions. Each dot represents the SHAP value of one patient, with a color gradient from purple to yellow indicating normalized feature value from low to high. The horizontal position shows the impact: negative values (left) lower disease risk and positive values (right) increase the risk. SHAP values on the x-axis represent the magnitude of impact on the heart disease risk prediction. ST slope shows a clear discriminatory pattern, with low values associated with strongly negative contributions and high values with strongly positive contributions to the predicted risk. This response illustrates how SHAP captures linear directional feature effects. Sex shows a bimodal clustering pattern, which shows that for categorical predictors with strong and consistent effects, their SHAP values tend to cluster by categories. Higher age values are associated with higher model-predicted risk, but its variable effect suggests interaction with other factors. Resting ECG and fasting blood sugar cluster near zero, which illustrates that SHAP can reveal variables that contribute minimally to risk prediction in the presence of other strong predictors. Oldpeak and maximum heart rate show dispersed, patient-specific effects, underscoring the value of patient-specific explanations.

**When to use:** SHAP Beewarm plot is a good tool to visualize the distribution of variable impact across variable values, rather than relying on a single number for average impact. This plot can show the magnitude, direction, and spread of effects for each variable across different patients.

**When not to use:** SHAP Beewarm can be more complex to interpret than SHAP waterfall plot. For specific patients' risk explanations, SHAP waterfall plot is a better option since it explicitly decomposes the predictions into additive contributions. Moreover, when there are certain clinically important subgroups with very few patients, SHAP Beewarm plot may be unsuitable, since the data points for rare but important categories may be overshadowed by the majority class in the summary plot, masking their clinical significance.

**Limitations:** Summary plots can mask patient-level details and feature interactions, particularly when the impact is bimodal or skewed. A feature with low average impact can be critical for specific subgroups, risking underestimation of its clinical importance. The color coding reflects only the raw feature values without clinical context: a "high" cholesterol might be normal or represent good control on statin therapy rather than risk.

### 3.2.2 LIME (Local Interpretable Model-Agnostic Explanations)

**What it is:** LIME [22] addresses the same patient-level question as SHAP - why did this patient receive this prediction - but uses a different approach. It fits a surrogate model in the neighborhood around a specific case. It generates perturbed samples around the patients' feature values, obtains the black-box model predictions for them, weights these samples by proximity, and then trains a sparse linear model (or some other interpretable models) whose coefficients approximate local feature importance.

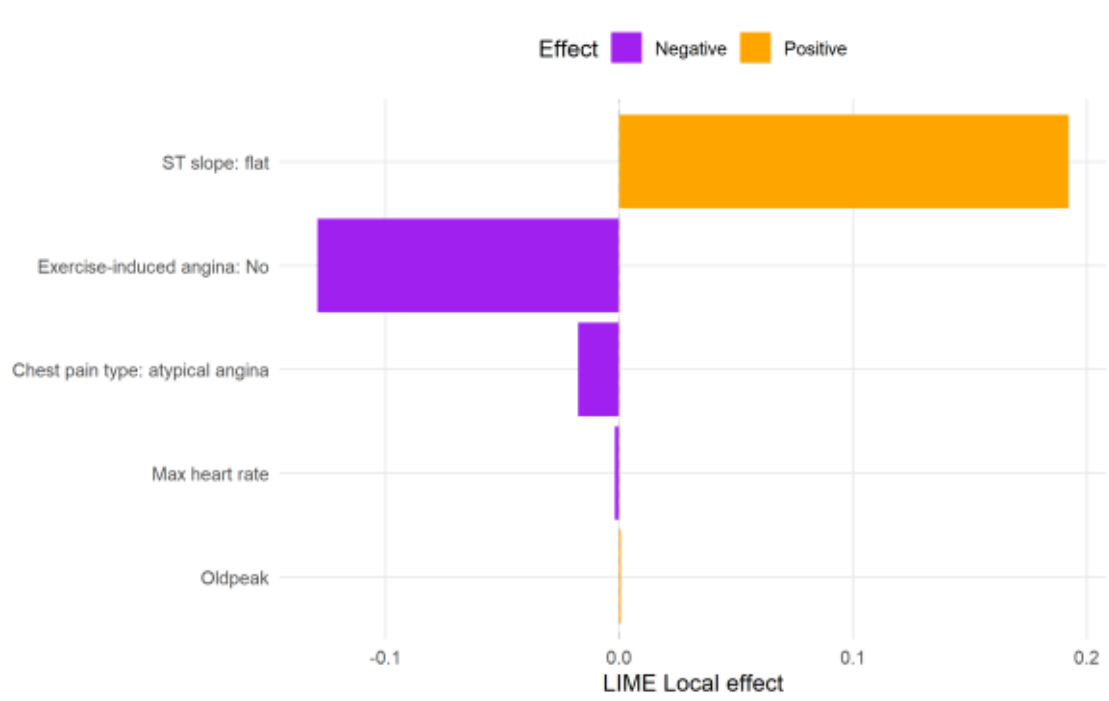

**Fig. 4.** LIME-style Local Explanation - Patient 22.

**Interpretation**: LIME (Fig. 4) explained Patient 22's low predicted heart disease risk by decomposing it into feature contributions: the absence of exercise-induced angina (-0.13) and chest pain type 2 (-0.02) contribute negatively to the predicted risk, while a flat ST slope (+0.19) contributes positively. Oldpeak contributes a small positive effect while maximum heart rate provides a small negative effect. For simplicity, the plot displays only the top 5 features obtained from sparse local linear approximation. This illustration shows how a complex model's local behavior around a single observation can be approximated, and individual predictions can be interpreted as a weighted sum of feature contributions.

**When to use:** LIME offers quick, straightforward, model-agnostic explanations by listing the most locally important features for a single prediction. It helps answer "why *this* prediction?" and is useful when Shapley values are too computationally expensive.

**When not to use:** LIME should not be used when we need consistent explanations for the overall predictions instead of specific cases, as LIME results can vary depending on how samples are generated. It can also be misleading when the simple local model does not accurately represent the true behavior of the underlying model.

**Limitations:** As a surrogate, LIME may misrepresent models with highly non-linear decision boundaries. Synthetic data generation introduces instability for patients in data-sparse regions or near decision boundaries. Small changes in local sampling or perturbation can produce different explanations for the same patient. The sparse feature selection might exclude clinically relevant factors. LIME's result depends on the neighborhood size and perturbation strategy; poor parameter selection may capture noise instead of meaningful model behavior. Unlike SHAP, LIME explanations are not additive and do not sum to the total prediction difference from baseline, making them difficult to verify for completeness.

**3.2.3. Partial Dependence Plots (PDP)**

**What it is:** Beyond individual patients, biomedical scientists often want to know how a single variable affects predicted risk across the full population. Partial dependence plots (PDP) [32] answer this by showing the marginal effect of features on predictions, averaging over the population. PDP answer: "If feature X was set to a specific value for everyone, what would the model predict on average?" Repeating this for a range of X values traces how the predictions change, offering a visualization for interactions and non-linearities effects. This allows biomedical scientists to verify expected patterns or identify counterintuitive trends. However, averaging may mask subgroup differences or interactions, which are better revealed by ICE plots.

**How to compute PD:** for a feature X (e.g.“age”): Set all individuals’ age to *x*, generate predictions with other features fixed, and average across all predictions. Repeating for a grid of *x* values (e.g. age = 20, 30, ..., 80) yields a curve (or a surface for two features).

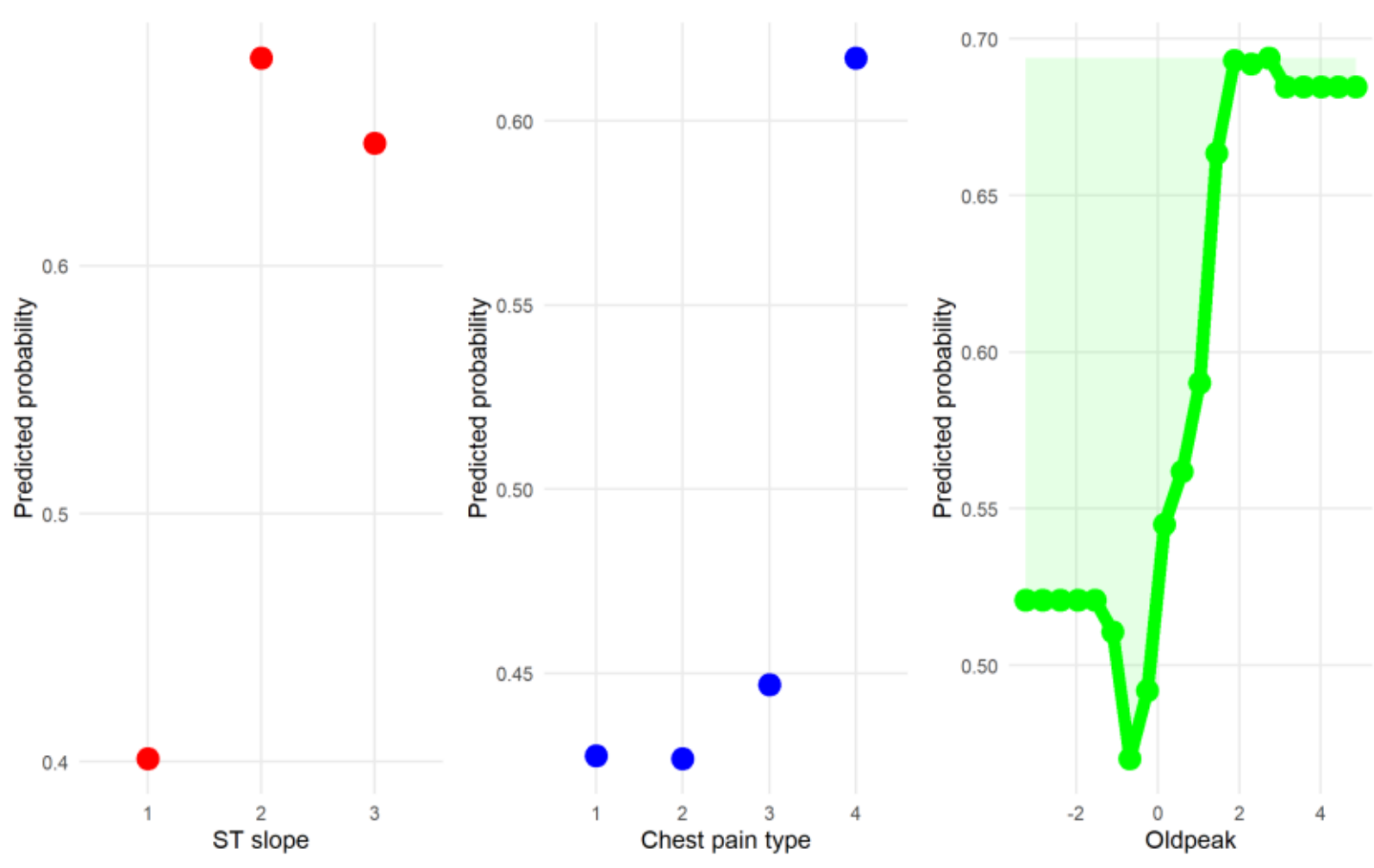

**Fig. 5.** Partial Dependence Plots for top 3 features.

**Interpretation**: PDP plots (Fig. 5) visualize features’ average effects on predicted probabilities across the top 3 predictors. **ST slope** shows a threshold effect: value 1 (upsloping) show low risk, flat and downsloping at values 2-3 show higher risk, with a slight decline at value 3. **Chest pain type** reveals a clearly distinct risk profile, with type 4 demonstrating substantially higher risk than types 1-3. Oldpeak shows a non-linear association with model-predicted risk: risk is lowest at -1, increases gradually with mild ST depression (-1-2mm), then steeply increases beyond 2mm, with a plateau at ~0.7 for severe depression (>3mm). These illustrates that PDP can reveal non-linear, step-wise, or diminishing marginal effects which are important to understand individual variable’s contribution to the risk predictions. These patterns in the model prediction are broadly consistent with commonly described diagnostic patterns in cardiology. However, this consistency does not constitute clinical validity. Whether these relationships generalize requires formal evaluation on a held-out test set.

**When not to use:** PDP should not be used when features are highly correlated, as it assumes feature independence. This may lead to visualization of impossible variable combinations and unrealistic relationships. PDP can also be misleading when strong interactions exist, since averaging can hide important patterns in the model; when datasets are small and not having enough points to average over reliably; when you need causal answers since PDP shows association, not cause-and-effect.

**Limitations:** PDPs can obscure heterogeneity and interaction effects by averaging across all instances. PDP also assumes feature independence, which is frequently invalid in clinical practice. For example, the impact of cholesterol might depend on age, blood pressure, and other factors, but PDPs cannot reveal these complex interdependencies. It is a good practice to complement PDPs with ICE plots or to plot PDPs stratified by a second feature (or use 2D PDP) to check for interaction effects.

### 3.2.4. Individual Conditional Expectation (ICE) Plots

**What it is:** Population averages can mask important patient-level variation. ICE [33] plots address this by showing individual prediction curves, revealing how changes in one feature affect specific cases. ICE plots reveal heterogeneity masked in population averages. They answer: for this patient, how does the model's predicted risk change as feature X isvaried while other features are held fixed". Parallel ICE lines indicate an additive, globally consistent effect (no major interactions) of the features; diverging or crossing ICE lines indicate interactions, where feature's effect depends on the patient's characteristics.

**How it's computed:** Similar to PDP: for each individual, vary the feature $X$ across a range, get model predictions, and plot the prediction curve. The PDP is the average of those curves.

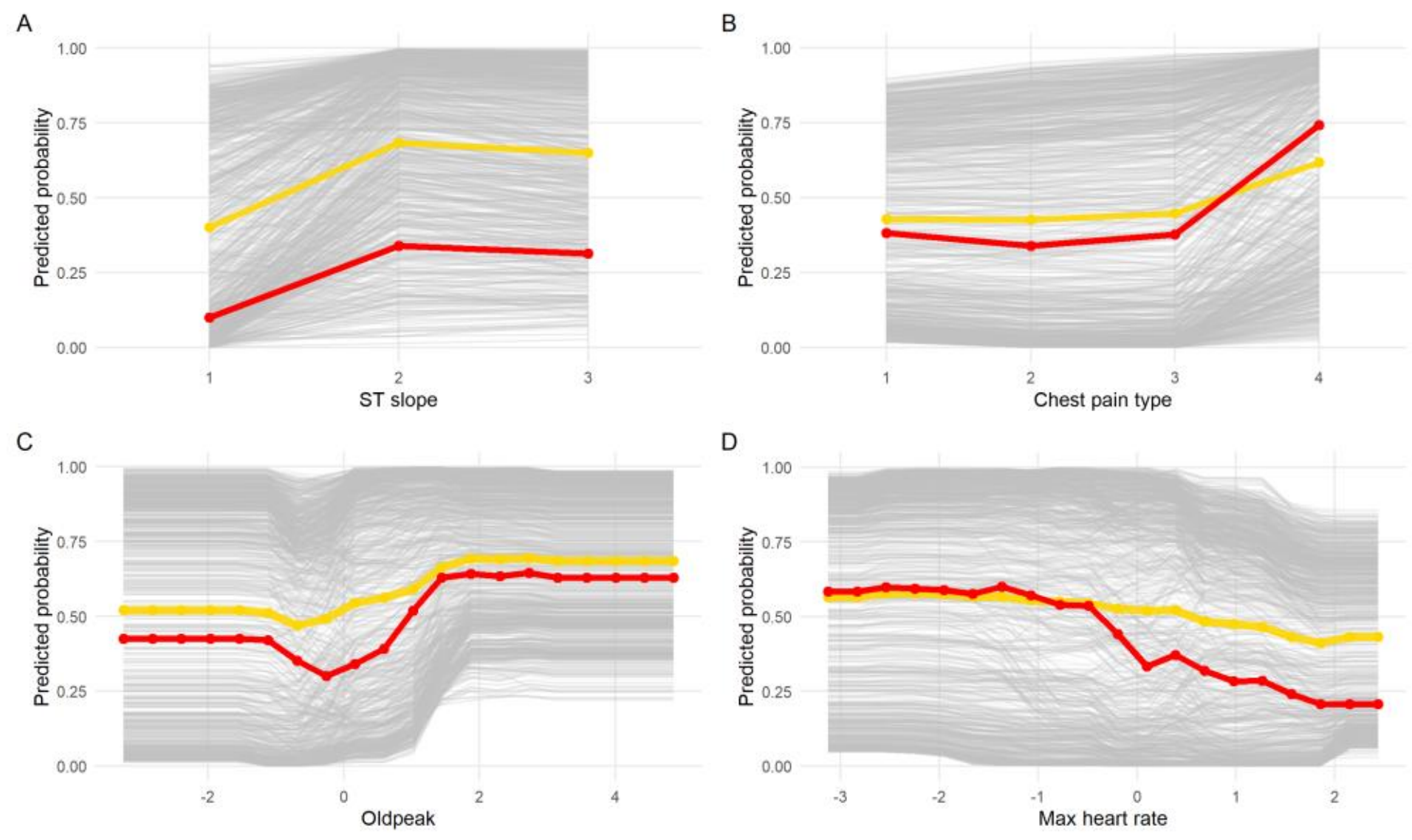

**Fig. 6.** ICE Plots for Top 4 Features. Patient 22 in red and PDP in yellow.

**Interpretation:** ICE plots (Fig. 6) visualize individual prediction trajectories across four key predictors. Each gray line corresponds to one patient, with Patient 22 highlighted in red, and the yellow line showing the population average PDP. **ST slope** shows heterogeneous trajectories between values 1 and 2, suggesting that the change from uploading to flat affects observations differently, and mostly parallel trajectories between values 2 and 3, suggesting a more globally consistent effect. Patient 22 follows the same directional trend as the population, while maintaining a consistently lower predicted risk than the population average across all ST slope values. **Chest pain type** exhibits heterogeneity and some patients have sharp risk increase between types 3 and 4, including Patient 22. **Oldpeak** displays even greater patient-specific variation. ICE curves diverge around 0: some are U-shaped, others monotonic. This crossing pattern is an illustration of how ICE plots can detect subgroup-level interaction effects that PDPs would smooth over. Higher **maximum heart rate** values are associated with lower model-predicted risk, with varying slopes across observations. The variation in trajectory gradients illustrates that ICE plots can quantify the degree to which a feature's effect is consistent versus individual-specific.

**When to use:** ICE plots are useful for precision medicine and model validation, confirming whether population-level effects apply to individuals. Divergence patterns in ICE highlight feature interactions that PDP alone will miss.

**When not to use:** ICE plots should not be used when you have too many instances, since the overlapping lines become unreadable clutter. They also should not be used when a clean global summary is needed.

**Limitations:** High-dimensional interactions are hard to display in ICE plots. Apparent heterogeneity can stem from noise, overfitting, or data limitations, rather than true clinical heterogeneity.

In summary, PDPs and ICE plots focus on marginal feature effects on model predictions. They show: "All else being equal, increasing X in the model input is associated with higher or lower model predictions on average". This is analogous to biomedical scientists examining a nomogram or risk calculator that shows the incremental risk associated with different levels of a factor. Interpretation should remain within the bounds of the observed data to avoid unreliable extrapolation.

## 3.3. Interpretable Models

### 3.3.1 Classical Interpretable Models

In high-stakes clinical applications, inherently interpretable models are often preferable to post-hoc explanations of black-box models, which may not reliably reflect model logic [42]. Interpretable models enable detection of biases in the training data and uncover discriminatory behavior in ML models [43]. Traditional models, such as logistic regression, rule-based models, and simple decision trees, are often considered more interpretable than complex black-box models because their prediction logic can be inspected directly.

**When to Use:** These should be the primary choice in high-stakes clinical settings where model transparency is a prerequisite for safety and regulatory compliance. They are ideal when the relationship between features is expected to be relatively linear or when a simple, deployable scoring system is required.

**When not to use:** Classical interpretable models should not be used when relationships are highly nonlinear or involve complex interactions, as they may oversimplify the data. They can also be misleading if their assumptions are violated, leading to poor model fit and incorrect conclusions.

**Limitations:** Classical models often struggle to capture complex, non-linear interactions common in high-dimensional healthcare data.

#### 3.3.2 Rule-Based Models

Rule-based models output human-readable decision rules or formulas instead of any computed quantities. These include decision trees, decision lists, and sparse rule sets. The advantage is that the model is self-explanatory – the logic by which it makes predictions can be directly inspected. Many clinical scoring systems and risk calculators fall into this category, either as simple scoring rules or flowcharts. RuleFit algorithm is an example of such model.

**RuleFit (Rule-based Ensemble)**

RuleFit was developed by Friedman and Popescu (2008) [34], which combines the predictive power of ensembles with interpretability. It first generates many candidate rules by training small decision trees, each path is a rule (e.g., “age > 70 and BP > 140 -> outcome”). It then performs a regularized linear regression to fit and assign weights to those binary rules (true/false flags). The result is a set of if-then rules with coefficients. For instance, in our heart disease example, RuleFit might generate a rule such as: *"If Age > 60 AND Chest Pain Type = 4 AND ST Slope = 2, then Risk +1.2"*. This allows a clinician to directly access the model's logic.

**When to Use:** RuleFit is particularly effective when biomedical scientists seek "if-then" logic that mirrors diagnostic flowcharts while maintaining the predictive power.

**When not to use:** Rule-based ensembles should not be used when the feature space is high-dimensional, since the number of rules explodes and interpretability is lost. They can also be misleading if the rules capture spurious patterns or do not generalize well to small changes in the data.

**Limitations:** The primary drawback is "rule overlap," where a single patient may trigger multiple, sometimes conflicting rules, making the ultimate prediction harder to trace than a simple decision tree.

In summary, RuleFit produces an interpretable model that expresses the outcome as a weighted sum of human-readable rules. It tells the biomedical scientists which combinations of features are informative and how they are combined. This can sometimes mirror the way biomedical scientists reason, but with data-driven thresholds and selections. Bayesian Rule Lists (BRL) [44] is another method that extends from this concept, to learn a sparse decision list – essentially a series of ordered if-else rules – using Bayesian inference to impose simplicity.

## 4. Conclusion and Discussion

Our work illustrates explainable machine learning methods through a single demonstration task based on a random forest model applied to structured clinical data. This choice was intentional to emphasize conceptual understanding of explanation techniques in a setting that closely mirrors many clinical risk prediction models. Readers can inspect and directly interpret XML outputs in terms of named, recognizable variables (e.g., ST slope, chest pain type). However, since the primary aim is XML methods demonstration, no test-train split was implemented, and readers should not draw clinical conclusions from these examples.

Modern healthcare machine learning increasingly involves high-dimensional modalities such as medical imaging, genomics, and multimodal EHR data. The explainability techniques investigated within our work are also widely used in these contexts, although their visualization and interpretation may differ. For example, the model-agnostic methods (SHAP, LIME, PDP, ICE) are broadly applicable to deep learning models as well. Future work could extend these demonstrations to multimodal or imaging-based models to further illustrate these techniques in contemporary clinical AI applications. Other XML methods specific to deep learning (eg: Saliency Maps, Layer-wise Relevance Propagation) that complement the model-agnostic approaches discussed here are reviewed in [45].

The complexity of ML models in clinical decision-making strengthens the need for explainable systems. XML offers biomedical scientists a toolkit to reason with models, connect them with clinical knowledge, and detect biases or unintended shortcuts in model logic. Explainability is key in model auditing and fairness diagnostics. However, no single explanation method suffices for all clinical contexts. Instead, biomedical scientists must adopt a toolbox approach: global methods (e.g. feature importance, PDPs) allow population-level validation for model predictions, while local methods (e.g. SHAP, LIME) support patient-level interpretability. Agreement between methods increases confidence that a feature is consistently influential in model predictions, while divergences highlight factors that need closer examinations.

Limitations specific to each XML method are discussed within their respective sections. Beyond these limitations, a common challenge across XML approaches is their potential to produce misleading explanations when input features are correlated or when the underlying model is misspecified, yielding outputs that may appear interpretable but lack reliability. Furthermore, the explanations generated by XML methods in this study should be interpreted as reflecting associations learned by the model, rather than causal relationships between variables and outcomes. Nonetheless, these methods remain valuable when applied with appropriate caveats, and mitigation strategies exist to address their limitations.

The field continues to evolve with several promising directions, including causal explanations ("What if this patient's blood pressure were lower?"), uncertainty quantification, and interactive explanation systems [46]. The ultimate goal of XML in healthcare is to foster a collaborative environment where biomedical scientists can actively shape model development, curate features, and contextualize predictions. The biostatistics community has a unique opportunity and responsibility to develop explanation methods that are both rigorous and usable, bridging the algorithmic sophistication of modern ML with the interpretive needs of clinical practice.

In summary, the integration of explainable ML into healthcare is not merely a technical challenge; it is a translational one. By grounding explanation methods in statistical intuition and presenting them in clinically resonant terms, we can foster a new generation of biomedical scientists capable of navigating and shaping the ML-enabled future of healthcare.

**CRediT Authorship Contribution Statement**

**Krishna Padmanabhan:** Conceptualization, Formal analysis, Methodology, Project administration, Supervision, Validation, Visualization, Writing – original draft, Writing – review & editing**. Minxin Lu:** Project administration, Validation, Visualization, Writing – original draft, Writing – review & editing. **Dai Feng:** Conceptualization, Writing – original draft, Writing – review & editing. **Natalia Kan-Dobrosky:** Conceptualization, Writing – original draft, Writing – review & editing. **Sai Konduri**: Writing – review & editing. **Heather J. Litman:** Conceptualization, Writing – original draft, Writing – review & editing. **Achilleas Livieratos:** Conceptualization, Writing – original draft, Writing – review & editing.

**Acknowledgements**

We thank Yunxun Wang for organizing and leading the working group. This collaboration would not have been possible without his support. AI-assisted language tools were used in a limited capacity for grammar checking, sentence concision, and stylistic editing, as well as for adding non-substantive comments to code. Scientific content, analyses, and interpretations were performed by the authors.

**Declaration of Competing Interest**

Heather J. Litman is an employee and stockholder of Thermo Fisher Scientific.

Dai Feng and Natalia Kan-Dobrosky are employees of AbbVie and own AbbVie stock.

All other authors declare no competing interests.

**Funding**

No external funding was received for this study.

**Data Availability**

The dataset used in this study is publicly available: Manu, S. (2024). *Heart Disease Dataset* [Data set]. Kaggle.

**References**

1 Alowais SA, Alghamdi SS, Alsuhebany N, *et al.* Revolutionizing healthcare: the role of artificial intelligence in clinical practice. *BMC Med Educ*. 2023;23:689.

2 Nilius H, Tsouka S, Nagler M, *et al.* Machine learning applications in precision medicine: Overcoming challenges and unlocking potential. *TrAC Trends Anal Chem*. 2024;179:117872.

3 Topol EJ. High-performance medicine: the convergence of human and artificial intelligence. *Nat Med*. 2019;25:44–56.

4 American Medical Association. AMA augmented intelligence research: Physician sentiments around the use of AI in health care—Motivations, opportunities, risks, and use cases: Shifts from 2023 to 2024. 2025. https://www.ama-assn.org/system/files/physician-ai-sentiment-report.pdf (accessed 4 December 2025)

5 Allgaier J, Mulansky L, Draelos RL, *et al.* How does the model make predictions? A systematic literature review on the explainability power of machine learning in healthcare. *Artif Intell Med*. 2023;143:102616.

6 Angelov PP, Soares EA, Jiang R, *et al.* Explainable artificial intelligence: an analytical review. *Wiley Interdiscip Rev Data Min Knowl Discov*. 2021;11:e1424.

7 Belle V, Papantonis I. Principles and practice of explainable machine learning. *Front Big Data*. 2021;4:688969.

8 Gilpin LH, Bau D, Yuan BZ, *et al.* Explaining explanations: An overview of interpretability of machine learning. *2018 IEEE 5th International Conference on data science and advanced analytics (DSAA)*. IEEE 2018:80–9.

9 Gunning D, Stefik M, Choi J, *et al.* XAI—Explainable artificial intelligence. *Sci Robot*. 2019;4:eaay7120.

10 Roscher R, Bohn B, Duarte MF, *et al.* Explainable machine learning for scientific insights and discoveries. *Ieee Access*. 2020;8:42200–16.

11 Molnar C. *Interpretable machine learning*. Lulu. com 2020.

12 Carvalho DV, Pereira EM, Cardoso JS. Machine learning interpretability: A survey on methods and metrics. *Electronics*. 2019;8:832.

13 Petch J, Di S, Nelson W. Opening the black box: the promise and limitations of explainable machine learning in cardiology. *Can J Cardiol*. 2022;38:204–13.

14 Doshi-Velez F, Kim B. Towards a rigorous science of interpretable machine learning. *ArXiv Prepr ArXiv170208608*. 2017.

15 Band SS, Yarahmadi A, Hsu C-C, *et al.* Application of explainable artificial intelligence in medical health: A systematic review of interpretability methods. *Inform Med Unlocked*. 2023;40:101286.

16 Gerlings J, Jensen MS, Shollo A. Explainable AI, but explainable to whom? An exploratory case study of xAI in healthcare. *Handbook of Artificial Intelligence in Healthcare: Vol 2: Practicalities and Prospects*. Springer 2021:169–98.

17 Lipton ZC. The mythos of model interpretability: In machine learning, the concept of interpretability is both important and slippery. *Queue*. 2018;16:31–57.

18 Srinivasu PN, Sandhya N, Jhaveri RH, *et al.* From blackbox to explainable AI in healthcare: existing tools and case studies. *Mob Inf Syst*. 2022;2022:8167821.

19 Zech JR, Badgeley MA, Liu M, *et al.* Variable generalization performance of a deep learning model to detect pneumonia in chest radiographs: a cross-sectional study. *PLoS Med*. 2018;15:e1002683.

20 Caruana R, Lou Y, Gehrke J, *et al.* Intelligible models for healthcare: Predicting pneumonia risk and hospital 30-day readmission. *Proceedings of the 21th ACM SIGKDD international conference on knowledge discovery and data mining*. 2015:1721–30.

21 Adamson AS, Smith A. Machine learning and health care disparities in dermatology. *JAMA Dermatol*. 2018;154:1247–8.

22 Ribeiro MT, Singh S, Guestrin C. " Why should i trust you?" Explaining the predictions of any classifier. *Proceedings of the 22nd ACM SIGKDD international conference on knowledge discovery and data mining*. 2016:1135–44.

23 U.S. Food and Drug Administration. Good Machine Learning Practice for Medical Device Development: Guiding Principles. 2021. https://www.fda.gov/media/153486/download (accessed 5 December 2025)

24 U.S. Food & Drug Administration. FDA Releases Artificial Intelligence/Machine Learning Action Plan. FDA. 2021. https://www.fda.gov/news-events/press-announcements/fda-releases-artificial-intelligencemachine-learning-action-plan (accessed 14 December 2025)

25 European Commission. EU Artificial Intelligence Act | Up-to-date developments and analyses of the EU AI Act. 2024. https://artificialintelligenceact.eu/ (accessed 4 December 2025)

26 U.S. Food and Drug Administration. Considerations for the Use of Artificial Intelligence To Support Regulatory Decision-Making for Drug and Biological Products. 2025. https://www.fda.gov/regulatory-information/search-fda-guidance-documents/considerations-use-artificial-intelligence-support-regulatory-decision-making-drug-and-biological (accessed 5 December 2025)

27 Ahmad MA, Eckert C, Teredesai A. Interpretable machine learning in healthcare. *Proceedings of the 2018 ACM international conference on bioinformatics, computational biology, and health informatics*. 2018:559–60.

28 Dwivedi R, Dave D, Naik H, *et al.* Explainable AI (XAI): Core ideas, techniques, and solutions. *ACM Comput Surv*. 2023;55:1–33.

29 Selvaraju RR, Cogswell M, Das A, *et al.* Grad-cam: Visual explanations from deep networks via gradient-based localization. *Proceedings of the IEEE international conference on computer vision*. 2017:618–26.

30 Montavon G, Binder A, Lapuschkin S, *et al.* Layer-wise relevance propagation: an overview. *Explain AI Interpret Explain Vis Deep Learn*. 2019;193–209.

31 Lundberg SM, Lee S-I. A unified approach to interpreting model predictions. *Adv Neural Inf Process Syst*. 2017;30.

32 Friedman JH. Greedy function approximation: a gradient boosting machine. *Ann Stat*. 2001;1189–232.

33 Goldstein A, Kapelner A, Bleich J, *et al.* Peeking inside the black box: Visualizing statistical learning with plots of individual conditional expectation. *J Comput Graph Stat*. 2015;24:44–65.

34 Friedman JH, Popescu BE. Predictive learning via rule ensembles. 2008.

35 Yang H, Rudin C, Seltzer M. Scalable Bayesian rule lists. *International conference on machine learning*. PMLR 2017:3921–30.

36 Siddhartha M. Heart Disease Dataset. 2024. https://www.kaggle.com/datasets/mexwell/heart-disease-dataset (accessed 4 December 2025)

37 American Heart Association. Coronary Artery Disease - Coronary Heart Disease. www.heart.org. 2024. https://www.heart.org/en/health-topics/consumer-healthcare/what-is-cardiovascular-disease/coronary-artery-disease (accessed 4 December 2025)

38 Birnbaum Y, Wilson JM, Fiol M, *et al.* ECG diagnosis and classification of acute coronary syndromes. *Ann Noninvasive Electrocardiol*. 2014;19:4–14.

39 Quinlan JR. Induction of decision trees. *Mach Learn*. 1986;1:81–106.

40 Breiman L. Random forests. *Mach Learn*. 2001;45:5–32.

41 Chen T. XGBoost: A Scalable Tree Boosting System. *Cornell Univ*. 2016.

42 Rudin C. Stop explaining black box machine learning models for high stakes decisions and use interpretable models instead. *Nat Mach Intell*. 2019;1:206–15.

43 Chen I, Johansson FD, Sontag D. Why is my classifier discriminatory? *Adv Neural Inf Process Syst*. 2018;31.

44 Letham B, Rudin C, McCormick TH, *et al.* Interpretable classifiers using rules and bayesian analysis: Building a better stroke prediction model. 2015.

45 Ras G, Xie N, Van Gerven M, *et al.* Explainable deep learning: A field guide for the uninitiated. *J Artif Intell Res*. 2022;73:329–96.

46 Raita Y, Camargo Jr CA, Liang L, *et al.* Big data, data science, and causal inference: a primer for clinicians. *Front Med*. 2021;8:678047.